# AI safety: state of the field through quantitative lens

Mislav Juric*, Agneza Sandic ** and Mario Brcic**
* Student at University of Zagreb Faculty of Electrical Engineering and Computing, Zagreb, Croatia
** University of Zagreb Faculty of Electrical Engineering and Computing, Zagreb, Croatia
mislav.juric2@fer.hr, agneza.sandic@fer.hr, mario.brcic@fer.hr

Abstract - Last decade has seen major improvements in the performance of artificial intelligence which has driven wide-spread applications. Unforeseen effects of such mass-adoption has put the notion of AI safety into the public eye. AI safety is a relatively new field of research focused on techniques for building AI beneficial for humans. While there exist survey papers for the field of AI safety, there is a lack of a quantitative look at the research being conducted. The quantitative aspect gives a data-driven insight about the emerging trends, knowledge gaps and potential areas for future research. In this paper, bibliometric analysis of the literature finds significant increase in research activity since 2015. Also, the field is so new that most of the technical issues are open, including: explainability and its long-term utility, and value alignment which we have identified as the most important long-term research topic. Equally, there is a severe lack of research into concrete policies regarding AI. As we expect AI to be the one of the main driving forces of changes, AI safety is the field under which we need to decide the direction of humanity's future.

## I. Introduction

The field of Artificial Intelligence (AI) safety is concerned with answering a very important question (along with its variations) - „How can we make AI safe for humans?". As the field of AI safety evolved over the years, many developments have been published on the topics related to AI safety. Although there are a lot of survey papers giving an overview of different aspects of the field of AI safety, there is a lack of quantitative insight into the state of the field of AI safety. Namely, there is a lack of data which tells a story about current and past trends and gives us empirical evidence to where research contributions would be most valuable.

In this paper we shall divide AI safety in the following hierarchy of sub-fields:

1. technical AI safety - deals with the technical issues of achieving safety and utility. It is subdivided by SRA classification [1]

    a) specification (S) – defines the purpose of the system

    b) robustness (R) – designing systems withstanding perturbations

    c) assurance (A) – monitoring, understanding, and controlling system during operation

2. AI ethics – mainly deals with the questions of moral responsibility and utility for the humans.

3. AI policy – deals with the questions how legal and governance systems need to be setup with respect to the new AI technologies (not only on national, but also on supranational level)

This paper is structured as follows. We shall explain our research methodology in section II, present our results and elaborate upon them in section III. In section IV we give our view on the current and possible future developments in the field of AI safety. Finally, in section V we shall give our conclusion and state the limitations to our research.

## II. Research methodology and scope

In this paper, we use bibliometrics as the basis of our research. In AI safety there are plenty of surveys (see Table 1), but most of them are specific to their certain sub-fields. There is a general survey [2], but it is of enumerative and descriptive nature, while we aim to supplement it with quantitative overview.

We have used online databases to identify indexed work sampled across different subfields of AI safety. We used the following databases: SCOPUS, Web of Science (WoS), and Google Scholar (GS). Since the area of AI safety is rather new, number of published work is smaller than for the well-established areas. This is especially the case for WoS. GS offers high-volume of work, including important pre-print sources, but the quality varies substantially and the result exploration seems to be limited. SCOPUS offers good trade-off between the volume and quality, paired with flexible searching and result exploration. So, we have manually fine-tuned our search-queries to SCOPUS and then we have used them across all three databases. We have fine-tuned queries until we got purity of 90% over returned results – purity established empirically by sampling across the results. We did our best to look into as many as subfields as possible, but we leave improvements, with respect to covered sub-fields, to the future work.

We have searched the following areas: AI ethics, AI policy, robustness (R), explainability and intepretability (A), fairness (S), value alignment (S), reward hacking (S), interruptibility (A), safe exploration (R), distributional shift (R), and AI privacy (A). Other areas such as verifiability have proven to be elusive for search queries in the terms of results' purity and relevance. The respective search queries for each selected area are given in the appendix VI.

Table 1 AI safety surveys

| Subject | References |
| --- | --- |
| General AI safety | [2]–[9] |
| AI ethics | [10]–[18] |
| AI policy | [19]–[24] |
| Interpretability/XAI | [25]–[32] |
| Adversarial robustness | [33]–[37] |
| Fairness/bias | [38]–[40] |
| Value alignment | [41]–[44] |
| Safe exploration | [45] |
| interruptibility | [46] |

Identified documents have the following distribution: conference papers (47.71%), journal articles (39.89%), reviews (3.89%), books (3.57%), book chapters (3.11%), and other (2.83%). We chose to focus on papers from 1985 till 2019 with the relevant data retrieved on January 26, 2020.

### III. AI SAFETY ARTICLES

In Figure 1 we can see the trends of growth in AI ethics and AI policy. AI ethics has seen steady growth since 2003, with visible explosion in interest since 2010. On the other hand, AI policy has had no significant amount of work until 2018 since when it is experiencing strong growth last two years.

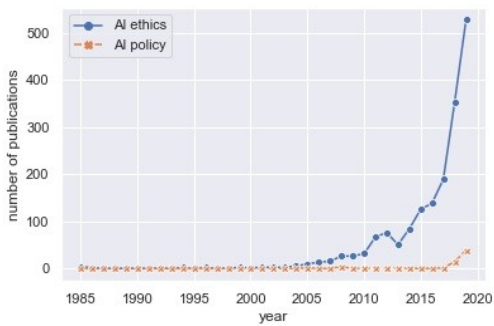

Figure 1 Number of papers in AI ethics and policy papers published each year

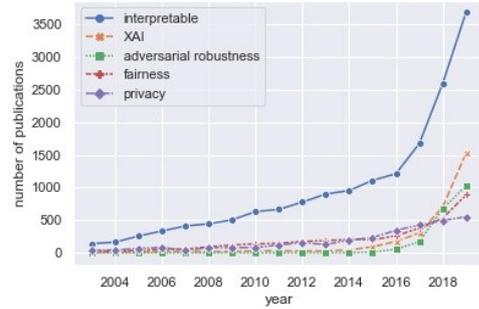

Figure 2 Number of papers in technical AI safety, high volume topics

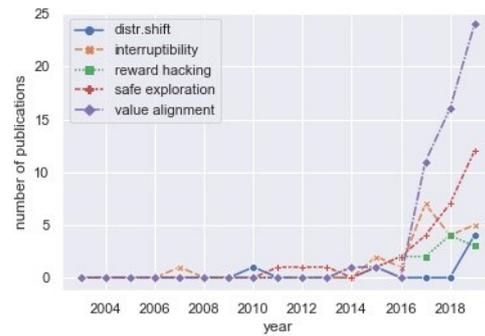

Figure 3 Number of papers in technical AI safety, low volume topics

Figure 2 and Figure 3 show the change of popularity for different subfields of technical AI safety over the last 20 years, grouped by their popularity level. The whole field of AI safety is seeing strong growth which is mainly driven by strong growing sub-fields. **Interpretability (with explainable AI (XAI) )** is the *single most* important growth generator for the field. *Strong growth* is shown by the fields of AI ethics, adversarial robustness and by the smaller volume topics of value alignment and safe exploration. These fields are driven mainly by the near-term applications in diversity of areas, such as transportation, medicine, biology, robotics, etc. Slight growth includes topics of fairness and privacy. Finally, the topics of safe exploration, distribution shift, interruptibility, and reward hacking seem to be emerging and are likely to become more intense venues in the future. Interruptibility and reward hacking are more focused on long-term research goals and do not get as much attention as other topics. These are areas that could be covered by non-publication decisions by some of the research organizations.

| rank | Journal | # | Journal | # |
| --- | --- | --- | --- | --- |
| 1 | Expert systems with applications | 249 | AI and Society | 32 |
| 2 | IEEE Access | 237 | Ethics and Information tech. | 32 |
| 3 | Neurocomputing | 200 | Futures | 17 |
| 4 | Information sciences | 199 | Minds and machines | 16 |
| 5 | Applied Soft Computing journal | 154 | Phil.and Tech./ Science and Engineering Ethics | 14 |

Table 2 Published articles per journal

In Table 2 we have identified top 5 journals for each technical AI safety and AI ethics respectively. We must emphasize that a considerable amount of interesting research and ideas in the area is not published in peer reviewed outlets. At best, they are published at pre-print services and otherwise they are published in blogs and their respective comments. This makes idea sharing harder and reinventing more likely. Some organizations, such as MIRI, have decided not to publish most of their work due to security reasons.

AI governance had such a small volume of work that journal ranking makes no sense. Some of the journals in AI policy are: „Computer Law And Security Review", „Contemporary Security Policy", and „Ethics And International Affairs".

| Tool/Technique | # |
|---|---|
| Machine Learning | 3052 |
| Classification | 2089 |
| Deep Learning | 1910 |
| Forecasting | 1289 |
| Feature Extraction | 1112 |
| Decision Trees | 1017 |
| Fuzzy Systems | 929 |
| Optimization | 899 |
| Clustering Algorithms | 763 |
| Genetic Algorithms | 738 |
| Knowledge Based Systems | 636 |

Table 3 Popular AI safety research tools and techniques

Table 3 summarizes most used tools and techniques in research based on their frequency as an article keyword. Machine learning tops the list, due to the latest surge of applications and proliferation. Classification seems to be the most studied problem setting, followed by forecasting. Deep learning is at the top of algorithmic approaches, followed by the decision trees which are more transparent model. Optimization is a tool used in learning systems and it can aim at different criteria. Genetic algorithms are an approach to optimization of hard problems. Fuzzy systems and knowledge based systems seem to be emerging in popularity, starting to be proposed as supplementary techniques to deep learning which bring complementary benefits where deep learning seems lacking.

IV. RECOMMENDATIONS FOR FUTURE RESEARCH

Generally, we welcome more concrete empirical work which would bring much needed information to fuel the future development. This calls for necessarily *multidisciplinary* work with both computational experimentation (such as [47]), and real-world testing (such as [48], [49]). Creation of many rich and diverse simulation environments and benchmarks would improve both system building and safety testing, since learning algorithms are data-hungry and real-world has too small experience bandwidth.

A. *Explainability*

One of the most important open problems in explainability is that there is no agreement on what an explanation is. Some works define decision tree, set of rules or an image as good explanation [31], with most of the work appealing to the mere intuition. Evaluation of comprehensibility of explanations to humans is underexplored [26]. There is still no algorithm that provides both high accuracy and explainability. In that regard, new hybrid techniques hold potential to achieve more effective explanations [32]. Looking further, the utility of explainability to the safety in the long-term is unclear since such approaches lack guarantees for adversarial schemes. Namely, possibly harmfully incorrect, but plausible explanations can be generated. Moreover, the size and dynamic of gap between the true and the most incorrect plausible explanation are interesting questions. Also, the nature of limits to the explainability of advanced concepts (for example in future science) to humans is not yet understood. Overstepping such limits makes explanations too far removed from the reality and their utility fades out.

B. *Value alignment*

Work on reinforcement learning (RL) and learning a reward function from actions and preferences will be very important for advancing the field of value alignment [2]. More advanced than value learning, value discovery through aligned algorithms could enable finding better reward functions that unlock new opportunities. It is more advanced prospect than value learning. Researching into reward corruption and side effects of optimizing for a goal which doesn't capture human values fully is just at the beginning. Currently, the most interesting work are approaches combining recursive factorization and bootstrapping for scalable reward learning and value alignment both in cooperative [50], [51] and adversarial [52] settings. There are many open questions about feasibility of such ideas with respect to error bounding, automated factorization process, validity of assumptions, etc. Mesa-optimizer [53] is an important concept, especially for the long-term research. It introduces multiple levels of value alignment within learned optimization. However, their relevance to the current practice is yet to be demonstrated.

C. *AI policy*

Transparency, accountability, reliability, security, corrigibility, interpretability, value specification, ability to limit capability, and performance and safety guarantees for particular AI systems are all important for future wide-spread applications and should be important parts and aims of future policies [21], [24]. Regulating research related to AI seems to be a possible long-term approach to regulating AI developments, but not at the cost of restricting scientific progress [5]. AI regulation could work globally only if there is concensus between the major AI research organizations. Although there are many papers related to AI governance, there is a severe lack of concrete AI policy suggestions.

### D. Corrigibility

Corrigibility, reasoning that reflects that an agent is incomplete and potentially flawed in a dangerous ways, is something that needs to be worked on [54]. Safety measures which are easily integrated within the AI development environment seems like something which would be tremendously valuable for improving AI safety. One promising method of improving AI safety in the short-term, while developing, is containment through virtualization [55].

### E. Safe exploration and distr. shift

Preventing catastrophic mistakes from occurring while training a reinforcement learning model is a non-negotiable need when the system interacts with the real world and not within a simulated environment [8]. Work in detecting and overriding an agent's action when it seems too dangerous seems like a good approach to reduce the number of catastrophic mistakes [2]. Proposing new conceptual solutions to the problem of safe exploration and distributional shift seems to be something that could bring a lot of value, but testing and/or extending existing solutions may be valuable as well. Hybridization of symbolic and sub-symbolic approaches might be a valid approach here.

### F. Adversarial robustness

Deep neural network models are highly vulnerable to adversarial attacks, which curtails their applications. There are special methods which reduce success rates of different types of adversarial attacks, but there are no general defensive methods successful against all attacks [34]. Open questions include why do the adversarial examples exist in the first place and why are they transferable [36]. Security verification of models to adversarial attacks is an open research challenge [35].

### G. AI ethics

Further work into human trust is necessary, especially from the aspect of human-AI interaction [13]. In applications, privacy concerns need to be taken seriously if we are to construct AI which is ethical [18], [56]. Ethics vary across the globe and evolve over time [12], [48], [49]. More research into human moral preferences is important to guide our development and policies.

## V. CONCLUSIONS AND LIMITATIONS OF THE RESEARCH

In this paper, we have surveyed the field of AI safety through quantiative lens. We have found various trends in the field of AI safety. We identified interpretability and explainability as strongest research topic in the near-term, but we have raised the question of utility in the long-term since such approaches lack guarantees for adversarial schemes [25]. Also, there are limits to explaining increasingly advanced concepts (for example, in future science) in a comprehensible way to humans. After overstepping such limits, the utility of explanations would fade out. Robustness-based topics are growing in importance with the incoming technical applications. AI policy is seeing its first greater contributions, but there is a severe lack of concrete policies. However, we would like to point out value alignment as the most important subfield in the long-term as it continues its sudden growth from relative obscurity. It is the most promising research direction for achieving the coexistence and cooperation with the computationally more capable agents than ourselves. If their goals are aligned with ours, they have no incentive to harm. In the case of misalignment, it is hard to have general guarantees through other approaches against incentivized, computationally superior agents.

This research has limitations, some of them are: underestimation/overestimation due to the search queries, non-publication bias of some research fields in peer-reviewed outlets, bias due to the increase of number of publishing outlets, reliance on the descriptive statistics to derive facts and insights about AI safety as we did not read all of the covered papers.

## VI. APPENDIX

The queries from the Table 4 were used for searching through all the fields. The only exception is the query for AI privacy which was used only on the fields: title, abstract, and keywords.

Table 4 Search queries used for bibliometric research

| Topic | Search query |
| --- | --- |
| explainability | ("artificial intelligence" OR "machine learning" OR "AI") AND "explainable" |
| interpretability | ("interpretable" OR "interpretability") AND "artificial intelligence" |
| <general> | "AI safety" OR "safe AI" |
| Adversarial robustness | "adversarial examples" |
| AI ethics | "AI ethics" OR "machine ethics" OR "friendly AI" OR "good AI" OR ("superintelligence" AND "risk") OR ("existential risk" AND "AI") |
| AI policy | "AI policy" OR "AI governance" |
| fairness | ( "AI" OR "algorithmic" ) AND ( "bias" OR "fairness" OR "discrimination") AND "ethics" |
| Interruptibility&corrigibility | "AI" AND ( "interruptibility" OR "corrigibility" ) AND "risk" |
| Safe exploration | "AI" AND "safe exploration" |
| Distributional shift | "Artificial intelligence" AND "distributional shift" AND ( "generalization" OR "safety" ) |
| Reward hacking | ( "reward hacking" OR "wireheading" OR "reward tampering" OR "reward gaming" ) AND "AI" |
| AI privacy* | "privacy" AND ("AI" OR "artificial intelligence") |
| Value alignment | ("value alignment" OR "alignment problem" ) AND "artificial" AND "ethics" |